\documentclass[journal,twoside,web]{ieeecolor2}
\usepackage{generic}
\usepackage{cite}
\usepackage{amsmath,amssymb,amsfonts}
\usepackage{graphicx}
\usepackage{textcomp}
\usepackage{booktabs}
\usepackage{multirow}
\usepackage{threeparttable}
\usepackage{array}
\usepackage{url}
\usepackage{stfloats}
\usepackage{algorithm}
\usepackage{algorithmic}

\newcommand{\INPUT}{\item[\textbf{Input:}]}
\newcommand{\OUTPUT}{\item[\textbf{Output:}]}

\def\BibTeX{{\rm B\kern-.05em{\sc i\kern-.025em b}\kern-.08em
    T\kern-.1667em\lower.7ex\hbox{E}\kern-.125emX}}

\begin{document}
	
\bstctlcite{IEEEexample:BSTcontrol}

\title{Temporal Out-of-Distribution Detection for Asynchronous Motor Imagery Brain-Computer Interfaces}

\author{Chenhao~Liu, Siyang~Li, Luofei~Tan, and Dongrui~Wu
	\thanks{Chenhao Liu, Siyang Li, and Dongrui Wu are with the School of Artificial Intelligence and Automation, Huazhong University of Science and Technology, Wuhan, China.}
	\thanks{Corresponding author: Dongrui Wu, e-mail: drwu@hust.edu.cn.}
}

\maketitle

\begin{abstract}
Real online brain--computer interfaces operate on continuous electroencephalography (EEG) streams, where users are usually at rest and enter motor-imagery task states only intermittently. EEG windows may also arise from OOD MI activity outside the predefined control set. Conventional closed-set motor-imagery classifiers tend to assign such inputs to ID classes, which can cause erroneous control. To address this issue, this paper proposes a two-stage EEG detection framework for asynchronous motor-imagery brain--computer interfaces. A sliding-window mechanism continuously monitors EEG signals. The first stage uses an EEGNet-based rest/task gate to determine whether the current window should enter the control-decision process. The second stage performs ID MI classification and out-of-distribution detection only for task-state samples. To improve OOD rejection, we further propose TempDens, which combines classification-output energy, deep-feature density, and temporal-consistency scores to characterize distributional deviation from output, feature, and temporal-dynamic perspectives. Experimental results show that the proposed method effectively supports task-state detection and OOD MI recognition in continuous EEG streams, outperforming multiple conventional OOD baselines. This study reframes online motor-imagery control as a hierarchical decision problem involving continuous monitoring, state discrimination, ID classification, and OOD rejection.
\end{abstract}


\section{Introduction}
\label{sec:introduction}
\IEEEPARstart{B}{rain--computer} interfaces (BCIs) establish a direct pathway between the brain and external devices, enabling communication and control without relying on peripheral nerves or muscles \cite{Graimann2010,wolpaw2002bci}. With advances in neural engineering, EEG processing, and intelligent algorithms, BCIs have shown promise in rehabilitation, assistive control, and human--machine interaction \cite{lotte2018review}. For real-world closed-loop systems, the key objective is not only to classify offline EEG samples, but also to infer task-related control states from continuous EEG streams reliably and translate them into safe control actions \cite{mason2000brain,ding2025eeg}.

Motor imagery (MI) \cite{pfurtscheller1997motor} BCI is one of the most representative paradigms in noninvasive BCI research. MI tasks can induce sensorimotor rhythm modulations similar to those elicited by actual movement, even without overt limb execution; therefore, they are widely used to build EEG-based active control systems \cite{pfurtscheller2001motor,schirrmeister2017deep,eegnet2018}.

Nevertheless, most MI-BCI studies still follow a synchronous protocol, where cue signals define trial boundaries and classification is performed on presegmented fixed-length epochs \cite{wu2024motor}. This setting reduces online control to closed-set classification over known EEG segments, such as left-hand, right-hand, feet, or tongue imagery. Although convenient for controlled evaluation, it assumes that the system knows when the user enters the task state and that every test brain state belongs to a class observed during training.

In real online BCI systems, EEG is observed as a continuous stream rather than as discrete trials with known boundaries. Because users are at rest for most of the time and enter MI states only occasionally, an asynchronous BCI must first determine whether the current window is rest or task before classifying the MI category. More importantly, online brain states are not limited to rest and predefined MI classes. Inner speech, spontaneous cognition, attention shifts, unmodeled imagery, or sudden responses to external events may produce task-state EEG patterns and pass the rest/task gate, although they do not belong to any ID control class. A closed-set MI classifier will nevertheless force such OOD samples into ID control commands, creating a direct safety risk in wheelchairs, robotic arms, cursors, or rehabilitation devices.

Out-of-distribution (OOD) \cite{yang2021oodsurvey} detection provides a principled way to identify samples that deviate from the distribution observed during training. Instead of forcing every input into a predefined class, an OOD-aware model estimates whether the current sample is sufficiently consistent with known data before issuing a prediction. This mechanism is well aligned with asynchronous MI-BCI, where task-state EEG activity may correspond either to an ID control class or to an OOD state that should be rejected.

This paper studies OOD rejection for asynchronous MI-BCI under continuous EEG streams. The main contributions are as follows. First, we formulate online MI control as a hierarchical decision process consisting of rest/task gating, in-distribution (ID) MI classification, and OOD rejection. Second, we propose TempDens, a temporal-density OOD score that integrates output energy, feature-space density, and short-term temporal consistency. Third, we evaluate the framework on public MI-EEG datasets by treating predefined MI categories as ID classes and held-out MI categories as OOD classes. The results show that temporal-density scoring improves task-state OOD detection over a broad set of post hoc baselines.

The remainder of this paper is organized as follows. Section~\ref{sec:related} reviews related work. Section~\ref{sec:method} presents the proposed method. Section~\ref{sec:experiments} describes the experiments and results. Section~\ref{sec:discussion} discusses findings and limitations. Section~\ref{sec:conclusion} concludes this paper.

\section{Related Work}
\label{sec:related}
\subsection{Motor-Imagery EEG Decoding}
MI-BCI decoding aims to infer imagined movement classes from EEG patterns over sensorimotor areas. During MI, event-related desynchronization and synchronization are commonly observed in the mu and beta rhythms,  providing a physiological basis for EEG-based MI decoding \cite{pfurtscheller1997motor,pfurtscheller2001motor}.Lotte \emph{et al.} \cite{lotte2018review} reviewed classical MI decoding pipelines based on band-pass filtering, spatial filtering, and shallow classifiers, where common spatial patterns are commonly used to extract class-discriminative spatial variance . Schirrmeister \emph{et al.} \cite{schirrmeister2017deep} proposed deep and shallow convolutional networks for end-to-end EEG decoding and demonstrated that convolutional architectures can learn informative temporal and spatial representations from raw EEG. Lawhern \emph{et al.} \cite{eegnet2018} introduced EEGNet, a compact convolutional architecture that combines temporal convolution, depthwise spatial filtering, and separable convolution for efficient EEG decoding. Because EEGNet is lightweight and suitable for online inference, it is adopted as the backbone in both stages of this work.

\subsection{Asynchronous BCI and State Gating}
Asynchronous BCI systems remove the assumption that trial timing is externally specified and require the system to decide when to respond. Mason and Birch \cite{mason2000brain} first introduced a brain-controlled switch for asynchronous control applications, emphasizing that practical BCI systems must continuously separate intentional control from noncontrol periods. Wu \emph{et al.} \cite{wu2024motor} proposed SWPC, a two-stage framework for asynchronous EEG-based MI classification that first detects task-related windows and then performs MI classification, highlighting the importance of reducing false activations in continuous EEG streams. Recent closed-loop demonstrations, such as the EEG-based real-time robotic hand control system of Ding \emph{et al.} \cite{ding2025eeg}, also show that online BCI deployment requires stable state detection before reliable device control can be achieved. In this work, asynchronous operation is addressed by a dedicated rest/task gate. However, rest/task discrimination alone is insufficient because a task-state input may still be outside the ID MI label space. The proposed second stage therefore performs OOD rejection within the task-state stream.

\subsection{Out-of-Distribution Detection}
OOD detection aims to identify inputs that deviate from the training distribution before a model produces an overconfident and potentially unsafe prediction. In a generalized OOD setting, distributional shifts may arise from semantic novelty, covariate changes, domain shifts, or open-set categories, and the detector must distinguish reliable in-distribution samples from inputs that should be rejected \cite{yang2021oodsurvey}. This problem is especially relevant to online BCI, where non-ID task-state EEG may still fall outside the predefined control vocabulary.

Beyond general image classification, OOD detection has been applied to safety-critical and open-world scenarios. In autonomous driving, anomaly segmentation benchmarks such as Fishyscapes \cite{blum2021fishyscapes} and SegmentMeIfYouCan \cite{chan2021segmentme} evaluate whether perception systems can detect unexpected objects. In medical image analysis, OOD detection is used to identify unseen scanners, acquisition protocols, patient populations, or novel diseases \cite{mohseni2021dermoscopic}. In natural language processing, out-of-scope intent detection benchmarks and methods help dialogue systems reject unsupported user intents \cite{larson2019clinc,zhan2021oosintent}. These applications share the same requirement as asynchronous MI-BCI: the model should know when not to make a forced decision.

Post hoc OOD methods have developed from confidence-based scores to logits, features, gradients, and activation-statistics-based detectors. MSP \cite{hendrycks2017baseline} uses the maximum softmax probability as a simple confidence score, and ODIN \cite{liang2018odin} improves softmax scoring with temperature scaling and input perturbation. Energy \cite{liu2020energy} replaces softmax confidence with a log-sum-exp energy score. OpenMax \cite{bendale2016openmax} models activation-vector tails for open-set recognition, while G-OpenMax \cite{ge2017gopenmax} extends OpenMax with generative modelling of unknown categories. GODIN \cite{hsu2020godin} further introduces a decomposed confidence formulation for generalized ODIN-style detection. Later methods exploit richer internal signals: GRAM \cite{sastry2020gram} uses Gram-matrix feature statistics, GradNorm \cite{huang2021gradnorm} measures gradient magnitudes, ReAct \cite{sun2021react} rectifies abnormal activations, DICE \cite{sun2022dice} suppresses less informative weights, and VIM \cite{wang2022vim} combines residual feature information with logits. Unlike these static single-window methods, TempDens explicitly incorporates short-term EEG feature dynamics.

\section{Method}
\label{sec:method}
\subsection{Problem Setup}
Assume that the in-distribution motor-imagery (MI) label space consists of $K$ task classes,
denoted by $\mathcal{Y}_{\mathrm{ID}}=\{1,\ldots,K\}$.
Training data are collected from $L$ subjects, where the $l$-th subject provides
$\mathcal{D}_{l}=\{(X_{l}^{i},y_{l}^{i})\}_{i=1}^{n_{l}}$.
Here, $X_{l}^{i}\in\mathbb{R}^{C\times T}$ denotes an EEG trial with $C$ channels and $T$ time samples,
and $y_{l}^{i}$ denotes its label. The training data include resting-state samples and samples from
the ID MI classes, whereas OOD MI classes are unavailable during training.

\begin{figure*}[t]
\centering
\includegraphics[width=0.9\textwidth]{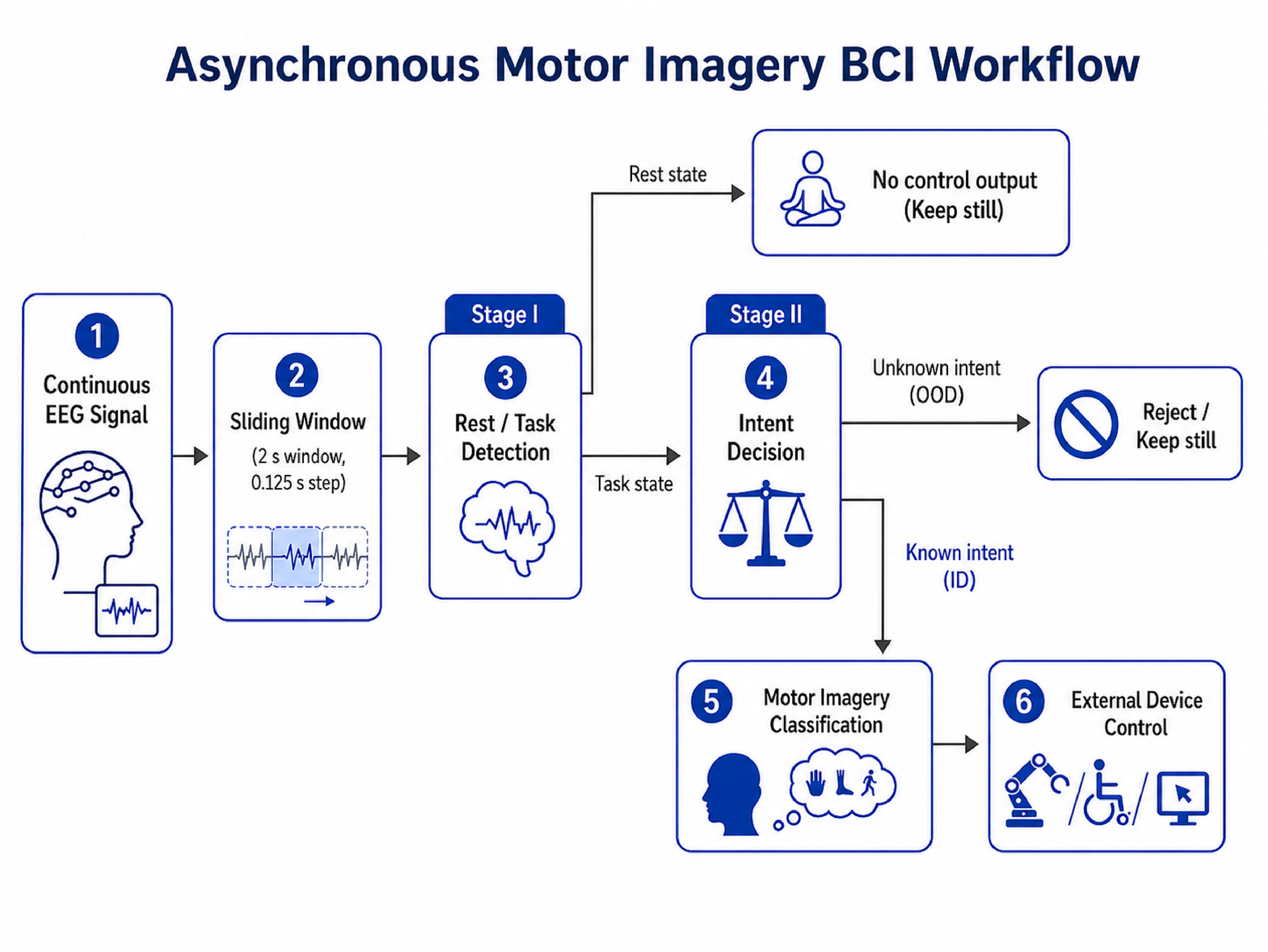}
\caption{Overall workflow of the proposed online asynchronous MI-BCI system. Continuous EEG is segmented into sliding windows, followed by rest/task gating, ID MI classification, and OOD rejection.}
\label{fig:overall_workflow}
\end{figure*}

At test time, unlabeled EEG windows from an unseen subject arrive sequentially.
At online step $a$, the decision for the current window $X_t^{a}$ can only rely on the training data
and the observed test stream $\{X_t^{i}\}_{i=1}^{a}$.
Each window is assumed to originate from one of three semantic states: resting state, ID MI activity, or OOD MI activity.
The online decision rule is defined as
\begin{equation}
	\pi(X_t^a)=
	\begin{cases}
		\mathrm{No\ Action}, & X_t^a \in \mathcal{D}_{\mathrm{rest}},\\
		\hat{y}_t \in \mathcal{Y}_{\mathrm{ID}}, & X_t^a \in \mathcal{D}_{\mathrm{ID}},\\
		\mathrm{Reject}, & X_t^a \in \mathcal{D}_{\mathrm{OOD}}.
	\end{cases}
\end{equation}
Thus, a control command is issued only when the current window is task-related and assigned to
one of the ID MI classes. As illustrated in Fig.~\ref{fig:overall_workflow}, this policy is implemented
by a two-stage framework consisting of rest/task gating followed by ID classification and OOD rejection,
without using labeled target-subject data.

\begin{figure*}[t]
\centering
\includegraphics[width=0.85\textwidth]{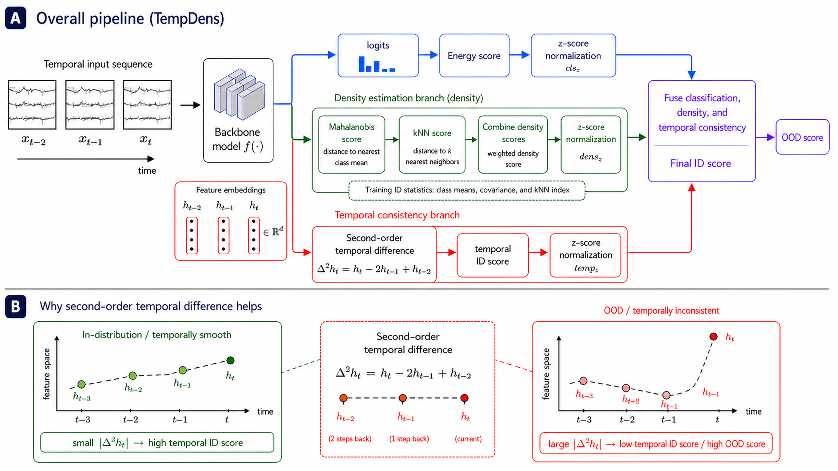}
\caption{Illustration of the proposed TempDens framework. (A) The TempDens pipeline, where task-state EEG windows are processed by ID MI classification and OOD scoring. (B) The rationale of the temporal-consistency term, showing how stable and unstable feature evolution helps distinguish ID task states from OOD task states.}
\label{fig:method}
\end{figure*}

\subsection{Stage I: Rest/Task Gating}
The first stage feeds each EEG window to an EEGNet binary classifier. The gate predicts whether the current window belongs to rest or task activity. Let $x_t\in\mathbb{R}^{C\times T}$ denote the current window, where $C$ is the number of EEG channels and $T$ is the number of temporal samples. The EEGNet gate outputs a task-state probability $p_{\mathrm{task}}(x_t)$, and the rest/task prediction is
\begin{equation}
\hat{s}_t=
\begin{cases}
\mathrm{task}, & p_{\mathrm{task}}(x_t)\geq \lambda,\\
\mathrm{rest}, & p_{\mathrm{task}}(x_t)<\lambda,
\end{cases}
\end{equation}
where $\lambda$ is the gating threshold. If the prediction is rest, the system directly outputs No Action and suppresses downstream decisions. To reduce ambiguity near trial boundaries, EEG segments within 0.5 s after MI offset are excluded from both training and evaluation. This exclusion avoids forcing transition windows with residual task-related activity into the rest class.

\subsection{Stage II: ID Classification and OOD Rejection}
When Stage I predicts task activity, the window is further processed by a second EEGNet model trained on the ID MI classes. This model outputs logits $z_t=[z_{t,1},\ldots,z_{t,K}]$ over $K$ ID MI classes and a latent feature vector $f_t=f(x_t)$ from the penultimate representation. A conventional closed-set system would directly assign the class label $\arg\max_k z_{t,k}$ to every task-state window, even when the window belongs to an OOD MI class. In contrast, the proposed framework computes an OOD score before issuing the command:
\begin{equation}
\hat{y}_t=
\begin{cases}
\mathrm{Reject}, & S_{\mathrm{OOD}}(x_t)>\tau,\\
\arg\max_k z_{t,k}, & S_{\mathrm{OOD}}(x_t)\leq\tau,
\end{cases}
\end{equation}
where $\tau$ is the rejection threshold. Thus, a control command is produced only when the task-state window is accepted as ID; otherwise, the system withholds control by outputting Reject.

\subsection{TempDens OOD Score}
TempDens rejects OOD task-state EEG windows by combining three complementary cues: classifier-output uncertainty, feature-space density, and temporal feature dynamics. The overall scoring pipeline and the motivation for the temporal term are shown in Fig.~\ref{fig:method}a.

The first term is an energy-based output score:
\begin{equation}
S_{\mathrm{ebo}}(x_t)=-T\log\sum_{k=1}^{K}\exp\left(z_{t,k}/T\right),
\label{eq:energy}
\end{equation}
where $T$ is the temperature. Higher values indicate weaker compatibility with the ID-class logit structure.

The second term is a feature-density score. Let $\mu_c$ be the mean feature vector of known class $c$ and let $\Sigma^{-1}$ be the inverse covariance estimated from training ID features. The nearest class-conditional Mahalanobis score is
\begin{equation}
S_{\mathrm{mahal}}(x_t)=\min_c (f_t-\mu_c)^\top \Sigma^{-1}(f_t-\mu_c).
\label{eq:mahal}
\end{equation}
To capture local density, an ID feature memory is used to compute the mean distance from $f_t$ to its $k$ nearest training features:
\begin{equation}
S_{\mathrm{knn}}(x_t)=\frac{1}{k}\sum_{i=1}^{k}d_i(f_t).
\label{eq:knn}
\end{equation}
The density score is then
\begin{equation}
S_{\mathrm{dens}}(x_t)=\eta S_{\mathrm{mahal}}(x_t)+(1-\eta)S_{\mathrm{knn}}(x_t),
\label{eq:density}
\end{equation}
where $\eta \in [0,1]$ balances global class geometry and local neighborhood density.

The third term is a temporal-consistency score. In continuous EEG streams, adjacent sliding windows are highly correlated because they contain overlapping signal segments and usually reflect the same ongoing mental state. Therefore, the features of ID samples are expected to vary smoothly once the MI task state becomes stable, since consecutive windows are drawn from the same learned class distribution and share substantial temporal content. In contrast, the features of OOD samples may change abruptly, because such samples are not constrained by the ID MI manifold and may contain nonstationary task-state patterns outside the predefined class space. As shown in Fig.~\ref{fig:method}b, this temporal instability provides evidence complementary to output confidence and feature density. We therefore use the second-order feature difference
\begin{equation}
S_{\mathrm{temp}}(x_t)=\left\|f_t-2f_{t-1}+f_{t-2}\right\|_2.
\label{eq:temp}
\end{equation}

Because the three scores have different scales, each score is standardized using the mean and standard deviation computed on training ID windows:
\begin{equation}
\widetilde{S}_{m}(x_t)=\frac{S_m(x_t)-\mu_m}{\sigma_m}, \quad
m\in\{\mathrm{ebo},\mathrm{dens},\mathrm{temp}\}.
\end{equation}
The final TempDens score is
\begin{equation}
S_{\mathrm{OOD}}(x_t)=
\alpha\widetilde{S}_{\mathrm{ebo}}(x_t)+
\beta\widetilde{S}_{\mathrm{dens}}(x_t)+
\gamma\widetilde{S}_{\mathrm{temp}}(x_t).
\label{eq:fusion}
\end{equation}

\begin{table}[t]
\centering
\caption{Mathematical Definitions of Distance Metrics and Temporal Features}
\label{tab:metrics_formulas}
\renewcommand{\arraystretch}{1.75}
\footnotesize
\begin{tabular}{lp{0.58\linewidth}}
\toprule
Metric & Formula \\
\midrule
Cosine & $d(i)=1-\frac{\sum_{j=1}^d h_{ij}c_j}{\sqrt{\sum_{j=1}^d h_{ij}^2}\sqrt{\sum_{j=1}^d c_j^2}}$ \\
Correlation & $d(i)=1-\frac{\sum_{j=1}^d (h_{ij}-\bar{h}_i)(c_j-\bar{c})}{\sqrt{\sum_{j=1}^d (h_{ij}-\bar{h}_i)^2}\sqrt{\sum_{j=1}^d (c_j-\bar{c})^2}}$ \\
Euclidean ($L_2$) & $d(i)=\sqrt{\sum_{j=1}^d (h_{ij}-c_j)^2}$ \\
Manhattan ($L_1$) & $d(i)=\sum_{j=1}^d |h_{ij}-c_j|$ \\
Canberra & $d(i)=\sum_{j=1}^d \frac{|h_{ij}-c_j|}{|h_{ij}|+|c_j|+\varepsilon}$ \\
Bray--Curtis & $d(i)=\frac{\sum_{j=1}^d |h_{ij}-c_j|}{\sum_{j=1}^d (|h_{ij}|+|c_j|)+\varepsilon}$ \\
Second-order Diff. & $d(i)=\|h_i-2h_{i-1}+h_{i-2}\|_2$\\

\bottomrule
\end{tabular}
\end{table}

\subsection{Overall Algorithmic Workflow}
The complete inference procedure follows the same hierarchical order as the proposed system design. First, the continuous EEG stream is filtered and segmented into overlapping 2-s windows with a 0.125-s update interval. Second, each window is passed to the Stage-I EEGNet gate. If the gate predicts rest, the system immediately outputs No Action and suppresses downstream control. Third, only task-state windows are passed to the Stage-II EEGNet classifier, which produces both logits and deep features. Fourth, TempDens computes output-energy, feature-density, and temporal-consistency scores, standardizes them using ID training statistics, and fuses them into a final OOD score. Finally, the system compares this score with the calibrated threshold: windows above the threshold are rejected as OOD samples, whereas windows below the threshold are assigned to the ID MI class with the largest logit response.

\begin{algorithm}[t]
\caption{Overall Online Decision Workflow of the Proposed Framework}
\label{alg:overall}
\begin{algorithmic}[1]
\INPUT \begin{tabular}[t]{@{}l@{}}
Continuous EEG stream $\mathcal{X}$;\\
Stage-I gate $G_{\theta}$;\\
Stage-II classifier $F_{\phi}$;\\
ID feature memory $\mathcal{M}$;\\
score statistics $\{\mu_m,\sigma_m\}$;\\
OOD threshold $\tau$.
\end{tabular}
\OUTPUT Online decision sequence $\{\hat{y}_t\}$
\STATE Initialize feature history queue $\mathcal{H}\leftarrow \emptyset$
\FOR{each update time $t$}
    \STATE Extract the latest 2-s window $x_t$ from $\mathcal{X}$
    \STATE $\hat{s}_t \leftarrow G_{\theta}(x_t)$ \COMMENT{Stage-I rest/task gating}
    \IF{$\hat{s}_t=\mathrm{rest}$}
        \STATE $\hat{y}_t \leftarrow \mathrm{No\ Action}$
        \STATE \textbf{continue}
    \ENDIF
    \STATE $(z_t,f_t)\leftarrow F_{\phi}(x_t)$ \COMMENT{Stage-II logits and features}
    \STATE Update $\mathcal{H}$ with $f_t$
    \STATE Compute $S_{\mathrm{ebo}}(x_t)$ from logits using \eqref{eq:energy}
    \STATE Compute $S_{\mathrm{dens}}(x_t)$ from $\mathcal{M}$ using \eqref{eq:mahal}--\eqref{eq:density}
    \STATE Compute $S_{\mathrm{temp}}(x_t)$ from recent features using \eqref{eq:temp}
    \STATE Standardize each score and compute $S_{\mathrm{OOD}}(x_t)$ using \eqref{eq:fusion}
    \IF{$S_{\mathrm{OOD}}(x_t)>\tau$}
        \STATE $\hat{y}_t \leftarrow \mathrm{Reject}$
    \ELSE
        \STATE $\hat{y}_t \leftarrow \arg\max_k z_{t,k}$
    \ENDIF
\ENDFOR
\end{algorithmic}
\end{algorithm}

\section{Experiments}
\label{sec:experiments}
\subsection{Datasets}
Experiments are conducted on two public MI EEG datasets. BNCI2014001 \cite{brunner2008bci}, also known as BCI Competition IV Dataset 2a, contains nine subjects, 22 EEG channels, 250-Hz sampling, and four MI classes: left hand, right hand, feet, and tongue. Following the cross-session setting, session 1 is used for training and session 2 for testing. Left- and right-hand imagery are used as ID classes, and feet and tongue imagery are used as OOD classes.

Zhou2016 \cite{zhou2016} contains four subjects, 14 EEG channels, 250-Hz sampling, and three MI classes: left hand, right hand, and feet. The first two sessions are used for training and the third session for testing. Left- and right-hand imagery are treated as ID, whereas feet imagery is treated as OOD. For both datasets, non-task intervals are used as rest-state samples after excluding ambiguous transition windows.

\begin{table*}[t]
\centering
\caption{SUMMARY OF THE TWO MI EEG DATASETS.}
\label{tab:dataset_summary}
\scriptsize
\setlength{\tabcolsep}{3pt}
\renewcommand{\arraystretch}{1.1}
\begin{tabular}{lccccccp{0.22\textwidth}p{0.15\textwidth}}
\toprule
Dataset & Subjects & Channels & Sampling Rate (Hz) & Trial Length (s) & Sessions & Trials & Types of Imagery & OOD Type \\
\midrule
BNCI2014001 & 9 & 22 & 250 & 4 & 2 & 144 & Left hand, right hand, feet, tongue & Feet and tongue imagery \\
Zhou2016 & 4 & 14 & 250 & 5 & 3 & 150 & Left hand, right hand, feet & Feet imagery \\
\bottomrule
\end{tabular}
\end{table*}

\subsection{Algorithms}
We compare TempDens with more than ten classical and state-of-the-art OOD detection algorithms, including confidence-based, logit-based, gradient-based, and feature-statistics-based approaches:
\begin{enumerate}
\item EEGNet \cite{eegnet2018}, a popular end-to-end convolutional neural network for EEG signal decoding. We use its version-4 implementation, which has two blocks of CNN layers and a fully connected classification layer.
\item Post hoc OOD approaches, including MSP, MaxLogit, ODIN, EBO, OpenMax, G-OpenMax, GradNorm, ReAct, DICE, VIM, and GRAM, all with EEGNet backbone.
\item Online OOD approaches, including online versions of the above post hoc detectors and TempDens, all with EEGNet backbone. Online methods additionally use recent sliding-window information through historical feature aggregation, so that the detector can exploit short-term temporal context.
\end{enumerate}

In the proposed framework, Stage I uses EEGNet as a binary rest/task gate, and Stage II uses EEGNet to produce ID-class logits and deep features for OOD scoring. The static OOD methods score the current task-state window independently, whereas the online OOD methods additionally use recent sliding-window information through historical feature aggregation to exploit short-term temporal context.

All experiments are conducted in a within-subject setting. For BNCI2014001, the first session is used for training and the second session for testing. For Zhou2016, the first two sessions are used for training and the third session for testing. Left- and right-hand imagery are treated as ID control classes, while the held-out MI classes listed in Table~\ref{tab:dataset_summary} are treated as OOD classes. Stage I is evaluated by rest/task accuracy and task recall under different task-state coverage ratios. Stage II OOD detection is evaluated by AUROC, where higher values indicate better ID/OOD separability.

EEGNet is trained with a batch size of 64 for both stages. Stage I and Stage II are each trained for 50 epochs using Adam with a learning rate of $10^{-3}$ and weight decay of $10^{-4}$. For TempDens, $\tau$ is calibrated on validation ID scores; unless otherwise stated, the temperature is set to $T=1$, $\eta=0.5$, and $\alpha=\beta=\gamma=1$.

All experiments are implemented in PyTorch, and the source code is available on GitHub\footnote{\url{https://github.com/Flashingcat/-Temporal-Out-of-Distribution-Detection.git}}.

\subsection{Rest/Task Gating}
\begin{figure}[t]
\centering
\includegraphics[width=\linewidth]{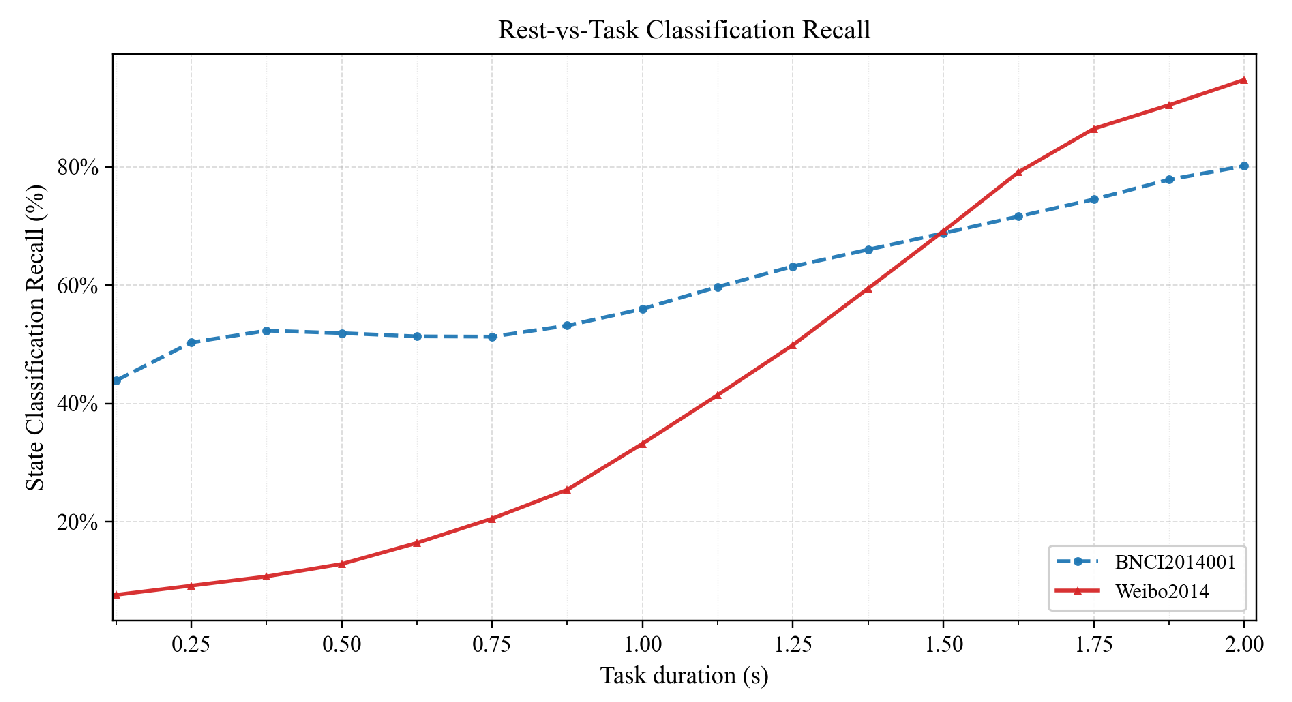}
\caption{Task-state recall as a function of task-state coverage in the 2-s sliding window. The horizontal axis denotes the proportion of MI activity within the window, and the vertical axis denotes the fraction of such windows detected as task/MI by the Stage-I gate.}
\label{fig:coverage}
\end{figure}

The Stage-I gate obtains accuracies of 0.8058 on BNCI2014001 and 0.7968 on Zhou2016. Fig.~\ref{fig:coverage} further analyzes gate behavior with respect to task-state coverage. The horizontal axis denotes the proportion of MI activity contained in a 2-s sliding window, i.e., the length of the MI segment divided by the full window length. The vertical axis denotes task recall, computed as the fraction of windows with a given coverage level that are correctly detected as task/MI windows by the Stage-I gate. The curve shows that task recall increases as the MI proportion within the window grows. This trend is expected: early after task onset, the window still contains substantial rest activity, whereas later windows contain stronger task-related EEG patterns. The result illustrates the latency--reliability tradeoff inherent to asynchronous sliding-window detection.

\subsection{OOD Detection Results}
Tables~\ref{tab:ood_bnci} and~\ref{tab:ood_zhou} report task-state OOD detection results on BNCI2014001 and Zhou2016, respectively. On BNCI2014001, static confidence-based methods, including EBO, MSP, MaxLogit, and ODIN, remain close to 0.50 average AUROC. Feature-structure methods perform better, with GRAM reaching 0.6461 in the static setting. However, GRAM degrades substantially in the online setting, indicating that simple temporal aggregation is not sufficient for robust EEG OOD detection. TempDens achieves the highest average AUROC of 0.6989 and is the best method on six of nine subjects.

On Zhou2016, the strongest conventional baseline is G-OpenMax with 0.6873 average AUROC in the online setting. TempDens improves the average AUROC to 0.8574 and obtains the best result on every subject. The improvement is especially clear for S1, S2, and S4, where temporal-density scoring substantially separates known left/right-hand imagery from the held-out feet imagery.

\begin{table*}[t]
\centering
\caption{OOD DETECTION AUROC ON BNCI2014001. THE BEST AUROC VALUES ARE MARKED IN BOLD, AND THE SECOND BEST BY AN UNDERLINE.}
\label{tab:ood_bnci}
\small
\setlength{\tabcolsep}{4pt}
\renewcommand{\arraystretch}{1.08}
\begin{threeparttable}
\scalebox{0.97}{\begin{tabular}{l|l|c|c|c|c|c|c|c|c|c|c}
\toprule
Setting & Approach & S1 & S2 & S3 & S4 & S5 & S6 & S7 & S8 & S9 & Average \\
\midrule
\multirow{11}{*}{Offline OOD}
& DICE      & 0.1860 & 0.5208 & 0.2083 & 0.3483 & 0.5358 & 0.4649 & 0.3228 & 0.4896 & 0.3021 & 0.3754 \\
& EBO       & 0.4426 & 0.5239 & 0.4910 & 0.4793 & 0.5932 & 0.4796 & 0.4171 & 0.6816 & 0.4037 & 0.5013 \\
& G-OpenMax & 0.5512 & 0.5031 & 0.6128 & 0.4872 & 0.5553 & 0.4915 & 0.4749 & \underline{0.7132} & 0.4602 & 0.5388 \\
& GRAM      & \underline{0.7576} & 0.4982 & \underline{0.8058} & \underline{0.6322} & 0.5038 & \underline{0.5416} & \textbf{0.7552} & 0.6711 & \underline{0.6491} & \underline{0.6461} \\
& GradNorm  & 0.2217 & \underline{0.5308} & 0.2992 & 0.4242 & 0.5578 & 0.4615 & 0.3546 & 0.5867 & 0.3421 & 0.4199 \\
& MSP       & 0.4211 & 0.5256 & 0.4959 & 0.4735 & 0.5928 & 0.4818 & 0.4313 & 0.6895 & 0.4195 & 0.5035 \\
& MaxLogit  & 0.4376 & 0.5245 & 0.4914 & 0.4769 & 0.5940 & 0.4798 & 0.4209 & 0.6832 & 0.4066 & 0.5017 \\
& ODIN      & 0.4239 & 0.5281 & 0.4742 & 0.4736 & 0.5607 & 0.4802 & 0.4209 & 0.6900 & 0.4176 & 0.4966 \\
& OpenMax   & 0.5759 & 0.4777 & 0.4759 & 0.5011 & 0.4525 & 0.4795 & 0.5266 & 0.6999 & 0.3984 & 0.5097 \\
& ReAct     & 0.5438 & 0.5189 & 0.5986 & 0.5046 & 0.5996 & 0.4892 & 0.4573 & 0.6977 & 0.4221 & 0.5369 \\
& VIM       & 0.6031 & 0.5052 & 0.6142 & 0.5606 & 0.5493 & 0.5185 & 0.5054 & 0.6559 & 0.5213 & 0.5593 \\
\midrule
\multirow{12}{*}{Online OOD}
& DICE      & 0.1945 & 0.5186 & 0.2199 & 0.3548 & 0.5378 & 0.4674 & 0.3295 & 0.4821 & 0.3125 & 0.3797 \\
& EBO       & 0.4379 & 0.5284 & 0.4871 & 0.4879 & 0.5976 & 0.4822 & 0.4175 & 0.6713 & 0.4000 & 0.5011 \\
& G-OpenMax & 0.5472 & 0.5121 & 0.5745 & 0.4891 & 0.5740 & 0.5044 & 0.4614 & 0.6908 & 0.4513 & 0.5339 \\
& GRAM      & 0.2454 & 0.5016 & 0.2686 & 0.3023 & 0.4794 & 0.4313 & 0.3789 & 0.4054 & 0.2858 & 0.3665 \\
& GradNorm  & 0.2440 & \textbf{0.5353} & 0.3216 & 0.4456 & 0.5770 & 0.4702 & 0.3612 & 0.6095 & 0.3503 & 0.4350 \\
& MSP       & 0.4182 & 0.5306 & 0.4920 & 0.4805 & 0.5967 & 0.4866 & 0.4324 & 0.6789 & 0.4140 & 0.5033 \\
& MaxLogit  & 0.4330 & 0.5289 & 0.4876 & 0.4844 & 0.5982 & 0.4839 & 0.4221 & 0.6732 & 0.4026 & 0.5016 \\
& ODIN      & 0.4184 & 0.5306 & 0.4920 & 0.4805 & 0.5967 & 0.4866 & 0.4324 & 0.6789 & 0.4140 & 0.5033 \\
& OpenMax   & 0.5539 & 0.4779 & 0.4247 & 0.4979 & 0.4428 & 0.4778 & 0.4775 & 0.6613 & 0.4216 & 0.4928 \\
& ReAct     & 0.5053 & 0.5208 & 0.5841 & 0.5070 & \underline{0.6003} & 0.4900 & 0.4580 & 0.6822 & 0.4062 & 0.5282 \\
& VIM       & 0.5329 & 0.5135 & 0.5615 & 0.5216 & 0.5737 & 0.4961 & 0.4668 & 0.6755 & 0.4913 & 0.5370 \\
& TempDens (ours) & \textbf{0.9195} & 0.5057 & \textbf{0.8662} & \textbf{0.6657} & \textbf{0.6272} & \textbf{0.5632} & \underline{0.6322} & \textbf{0.7834} & \textbf{0.7272} & \textbf{0.6989} \\
\bottomrule
\end{tabular}}
\end{threeparttable}
\end{table*}

\begin{table*}[t]
\centering
\caption{OOD DETECTION AUROC ON ZHOU2016. THE BEST AUROC VALUES ARE MARKED IN BOLD, AND THE SECOND BEST BY AN UNDERLINE.}
\label{tab:ood_zhou}
\small
\setlength{\tabcolsep}{8pt}
\begin{threeparttable}
\scalebox{0.97}{\begin{tabular}{l|l|c|c|c|c|c}
\toprule
Setting & Approach & S1 & S2 & S3 & S4 & Average \\
\midrule
\multirow{11}{*}{Offline OOD}
& DICE & 0.2325 & 0.0879 & 0.4466 & 0.1966 & 0.2409 \\
& EBO & 0.5395 & 0.3779 & 0.6771 & 0.5557 & 0.5375 \\
& G-OpenMax & 0.6794 & 0.6093 & \underline{0.6862} & 0.6825 & 0.6644 \\
& GRAM & \underline{0.8208} & \underline{0.7333} & 0.5029 & 0.6202 & 0.6693 \\
& GradNorm & 0.2446 & 0.2049 & 0.6029 & 0.2901 & 0.3356 \\
& MSP & 0.5625 & 0.4172 & 0.6788 & 0.5852 & 0.5609 \\
& MaxLogit & 0.5436 & 0.3894 & 0.6776 & 0.5638 & 0.5436 \\
& ODIN & 0.5472 & 0.4100 & 0.6739 & 0.5733 & 0.5511 \\
& OpenMax & 0.5012 & 0.5888 & 0.5331 & 0.5347 & 0.5395 \\
& ReAct & 0.7322 & 0.5243 & 0.6976 & 0.6224 & 0.6441 \\
& VIM & 0.6264 & 0.5216 & 0.6384 & 0.6383 & 0.6062 \\
\midrule
\multirow{12}{*}{Online OOD}
& DICE & 0.2417 & 0.0980 & 0.4445 & 0.2007 & 0.2462 \\
& EBO & 0.5343 & 0.3848 & 0.6687 & 0.5527 & 0.5351 \\
& G-OpenMax & 0.6806 & 0.6651 & 0.6840 & \underline{0.7195} & \underline{0.6873} \\
& GRAM & 0.2113 & 0.1837 & 0.4988 & 0.2167 & 0.2776 \\
& GradNorm & 0.2515 & 0.2041 & 0.6211 & 0.2932 & 0.3425 \\
& MSP & 0.5578 & 0.4261 & 0.6701 & 0.5818 & 0.5590 \\
& MaxLogit & 0.5391 & 0.3981 & 0.6692 & 0.5611 & 0.5419 \\
& ODIN & 0.5586 & 0.4261 & 0.6704 & 0.5818 & 0.5592 \\
& OpenMax & 0.4630 & 0.5828 & 0.5231 & 0.5330 & 0.5255 \\
& ReAct & 0.7235 & 0.5146 & 0.6802 & 0.6021 & 0.6301 \\
& VIM & 0.6017 & 0.4956 & 0.6690 & 0.6436 & 0.6025 \\
& TempDens (ours) & \textbf{0.8911} & \textbf{0.9431} & \textbf{0.7005} & \textbf{0.8949} & \textbf{0.8574} \\
\bottomrule
\end{tabular}}
\end{threeparttable}
\end{table*}

\subsection{Ablation Study}
Table~\ref{tab:ablation} analyzes the contribution of the three TempDens components. Energy alone is weak, with an average AUROC of 0.5194. Density alone is the strongest single component, reaching 0.7521, which indicates that latent feature geometry is highly informative for EEG OOD detection. Temp alone also performs well, reaching 0.7320. Combining density and temporal consistency without energy gives 0.7517, whereas the full model gives the best average AUROC of 0.7781 in this ablation setting. These results show that the three scores provide complementary evidence.

\begin{table}[t]
\centering
\caption{ABLATION STUDY RESULTS IN TERMS OF AUROC. THE BEST AUROC VALUES ARE MARKED IN BOLD.}
\label{tab:ablation}
\small
\setlength{\tabcolsep}{6pt}
\renewcommand{\arraystretch}{1.1}
\scalebox{0.94}{
\begin{tabular}{c|c|c|c|c|c}
\toprule
EBO & Dense & Temp & BNCI2014001 & Zhou2016 & Average \\
\midrule
$\checkmark$ & $\times$     & $\times$     & 0.5013 & 0.5375 & 0.5194 \\
$\times$     & $\checkmark$ & $\times$     & 0.6904 & 0.8139 & 0.7521 \\
$\times$     & $\times$     & $\checkmark$ & 0.6772 & 0.7868 & 0.7320 \\
$\checkmark$ & $\checkmark$ & $\times$     & 0.6531 & 0.7744 & 0.7138 \\
$\checkmark$ & $\times$     & $\checkmark$ & 0.6432 & 0.8393 & 0.7413 \\
$\times$     & $\checkmark$ & $\checkmark$ & 0.6982 & 0.8052 & 0.7517 \\
$\checkmark$ & $\checkmark$ & $\checkmark$ & \textbf{0.6989} & \textbf{0.8573} & \textbf{0.7781} \\
\bottomrule
\end{tabular}}
\end{table}

\subsection{Temporal Distance Metrics}
Table~\ref{tab:distance_metrics} compares distance metrics used for temporal discrepancy modeling. Manhattan distance gives the highest average AUROC and the best Zhou2016 result, whereas second-order $L_2$ gives the best BNCI2014001 result and a closely comparable average. This suggests that temporal discrepancy is useful across metric choices, but the optimal metric may depend on dataset-specific feature geometry and noise characteristics.

\begin{table}[t]
\centering
\caption{PERFORMANCE OF DIFFERENT DISTANCE METRICS UNDER TEMPDENS IN TERMS OF OOD AUROC. THE BEST AUROC VALUES ARE MARKED IN BOLD.}
\label{tab:distance_metrics}
\small
\begin{tabular}{lccc}
\toprule
Metric & BNCI2014001 & Zhou2016 & Average \\
\midrule
Bray--Curtis       & 0.6264 & 0.6420 & 0.6342 \\
Canberra           & 0.6352 & 0.6661 & 0.65070 \\
Correlation        & 0.6331 & 0.7056 & 0.6694 \\
Cosine             & 0.6183 & 0.6128 & 0.6155 \\
Euclidean          & 0.6945 & 0.8592 & 0.7769 \\
Manhattan          & 0.6955 & \textbf{0.8630} & \textbf{0.7792} \\
Second-order $L_2$ & \textbf{0.6989} & 0.8574 & 0.7782 \\
\bottomrule
\end{tabular}
\end{table}

\section{Conclusion and Future Work}
\label{sec:conclusion}

This paper presented a two-stage asynchronous MI-BCI framework with explicit OOD rejection for continuous EEG data streams. The first stage serves as a rest/task gate, preventing most non-control EEG windows from triggering unnecessary responses. The second stage is activated only for task-state windows and performs ID MI classification together with OOD rejection. This design reflects an important distinction in online BCI control: OOD task-state samples should not be simply merged with rest, because they may contain neural activations that resemble task activity while not belonging to any predefined control class. To address this problem, the proposed TempDens score combines energy-based output evidence, feature-space density, and temporal consistency, allowing the system to assess whether a task-state window belongs to the ID control set. Experiments on BNCI2014001 and Zhou2016 show that the proposed method improves OOD AUROC over multiple post hoc baselines, suggesting that safe online MI-BCI deployment requires not only accurate ID classification but also explicit mechanisms for deciding when the system should withhold control.

Several limitations remain, and future work will focus on the following directions:

\begin{itemize}
	\item \textbf{Real online closed-loop validation.} The current experiments replay public offline datasets as continuous streams. Future work should evaluate the framework in genuine online BCI control scenarios, where feedback adaptation, user fatigue, latency constraints, and behavioral changes may affect system performance.
	
	\item \textbf{Broader OOD-state taxonomy.} The current OOD samples are constructed from held-out MI categories, which represent controlled OOD classes. Future studies should include more realistic OOD states, such as unrelated cognitive activities, attention shifts, emotional fluctuations, artifacts, and unexpected external stimuli.
	
	\item \textbf{Lightweight real-time platform development.} The current framework has mainly been evaluated in an offline replay setting. Future work should build a lightweight real-time BCI platform that integrates EEG acquisition, sliding-window processing, model inference, OOD rejection, and external device control, so as to evaluate the end-to-end latency, computational efficiency, and practical deployability of the proposed system.
	
	\item \textbf{Multi-scale temporal modeling.} The current method uses a fixed window length and short-term temporal consistency. Future work may investigate multi-scale windows, temporal neural networks, or adaptive window mechanisms to better balance response latency and detection reliability.
	
	\item \textbf{Long-term robustness and deployability.} Future research should further examine the stability of the proposed framework under long-term use, cross-session distribution shifts, and real-world deployment conditions.
\end{itemize}

\bibliographystyle{IEEEtran}
\bibliography{references}

\end{document}